\documentclass[journal,twoside,web]{ieeecolor}

\usepackage{generic}
\usepackage{lcsys}
\usepackage{amsmath,amssymb,amsfonts,bm, mathtools, dsfont}
\usepackage{color}
\usepackage{cite}
\usepackage{textcomp}
\usepackage{xcolor}
\let\IEEEproof\proof
\let\IEEEendproof\endproof

\let\proof\relax
\let\endproof\relax

\usepackage{amsthm}

\let\proof\IEEEproof
\let\endproof\IEEEendproof
\usepackage{multirow}
\usepackage{tabularray}
\usepackage{tikz}
\usepackage{booktabs}
\usepackage{siunitx}
\usepackage{pgfplots}
\usepackage{pgfplotstable}
\usepgfplotslibrary{fillbetween}
\pgfplotsset{compat=1.18}
\usepackage{float}
\usepackage[caption=false]{subfig}
\usepackage{tikz}
\usetikzlibrary{matrix, positioning, intersections}
\usepackage{stmaryrd}

\newtheoremstyle{boldtheorem}
  {3pt}   
  {3pt}   
  {}      
  {}      
  {\bfseries}  
  {.}     
  {1em}   
  {}      

\newtheorem{definition}{Definition}
\newtheorem{theorem}{Theorem}
\newtheorem{lemma}{Lemma}
\newtheorem{proposition}{Proposition}

\newtheorem{assumption}{Assumption}

\usepackage{hyperref}
\usepackage{scalerel}
\usepackage{algorithm}   
\usepackage{algorithmic} 
\newcommand{\estl}{pacSTL}

\title{
pacSTL: PAC-Bounded Signal Temporal Logic from Data-Driven Reachability Analysis
}

\author{Hanna Krasowski$^{\ast, 1}$, Elizabeth Dietrich$^{\ast, 1}$, Emir Cem Gezer$^{2}$, \\ Roger Skjetne$^{2}$, Asgeir Johan Sørensen$^{2}$, and Murat Arcak$^{1}$ 
\thanks{This work was funded in part by the National Science Foundation (NSF) under Grant CNS-2111688, the Research Council of Norway under SFI AutoShip Grant 309230, and NTNU VISTA CAROS. Elizabeth Dietrich was also supported by the NSF GRFP.}
\thanks{$^{\ast}$Equal contribution}
\thanks{$^{1}$Elizabeth Dietrich, Hanna Krasowski, Murat Arcak are with the University of California, Berkeley, Berkeley, CA 94720 USA
         {\tt\small \{eadietri, krasowski, arcak\}@berkeley.edu}}%
 \thanks{$^{2}$Emir Cem Gezer, Roger Skjetne, Asgeir Sørensen are with the Norwegian University of Science and Technology, Trondheim, 7052 Norway
         {\tt\small \{emir.cem.gezer, roger.skjetne, asgeir.sorensen\}@ntnu.no}}%
}

\begin{document}
\maketitle
\thispagestyle{empty}

\begin{abstract}

Signal Temporal Logic (STL) is an expressive language for specifying behaviors of dynamical systems from continuous signals. However, a limitation of standard STL is its inherently deterministic semantics, which prevents it from accommodating uncertainty. 
Existing approaches to overcome this limitation are computationally costly and limit real-time capability, requiring repeated trajectory sampling or redesign of probability distributions over atomic propositions whenever the atomic propositions or specifications change. 
We introduce pacSTL, a framework that combines Probably Approximately Correct (PAC)-bounded reachable set predictions with an interval extension of STL. pacSTL computes lower and upper bounds on atomic robustness values by solving optimization problems over PAC-bounded reachable sets and propagates the bounds through the temporal logic operators. The resulting evaluation yields a PAC-bounded robustness interval at the specification level.
We demonstrate the efficiency and relevance of pacSTL by verifying a quadrotor flight scenario and runtime monitoring a maritime navigation encounter.

\end{abstract}



\section{Introduction}
\IEEEPARstart{S}{ignal} Temporal Logic (STL) is a formal specification language that encodes desired behaviors of dynamical systems for monitoring or verification \cite{maler2004-stllanguage, Fainekos2009}. 
It is increasingly used in robotic systems 
\cite{choe2025seeingsayingsolvingllmtotl, chen2024autotampautoregressivetaskmotion}, control synthesis \cite{YuDimarogonas-2024-STLcontrolsynthesis-TRO, SrinivasanCoogan2021-TL2CBF-TRO}, 
and reinforcement learning \cite{Li2017, Jiang_Bharadwaj_Wu_Shah_Topcu_Stone_2021}. Despite its versatility, a major limitation of standard STL is its inherently deterministic semantics, which prevent it from accommodating uncertainty. 
Recent work has explored three directions to address this limitation: set-based temporal logic formulations, probabilistic STL languages based on stochastic atomic propositions, and the integration of  statistical verification with STL robustness. 

For bounded uncertainty, set-based temporal logic languages have been proposed \cite{deshmukh2017robust, Baird2023,Roehm2016, KOCHDUMPER2024-TLverification}. 
For instance, the semantics in~\cite{deshmukh2017robust} robustly monitor partial signals using interval arithmetic. Recently, Interval-STL (I-STL) was introduced to reason about interval-bounded signals using inclusion functions, yielding robustness intervals for STL specifications with three-valued semantics (satisfaction, falsification, and undefined)~\cite{Baird2023}. 
Reachability-based temporal logics were introduced in~\cite{Roehm2016, KOCHDUMPER2024-TLverification} to enable verification of systems 
using reachable sets represented as polytopes or zonotopes. However, all of these temporal logic formulations either require linear atomic propositions \cite{Roehm2016, KOCHDUMPER2024-TLverification}, 
limiting specification expressiveness, or rely on interval approximations \cite{Baird2023}, potentially introducing significant conservativeness.

Probabilistic STL languages typically define atomic propositions that are satisfied with
a user-specified probability~\cite{sadigh2016safe, TIGER2020325, yoo2015control}. 
However, existing formulations rely on explicit probability distributions for atomic propositions \cite{sadigh2016safe, TIGER2020325}, or functions that estimate satisfaction probability for atomic propositions~\cite{yoo2015control}. Thus, these methods impose distributional assumptions and require effort whenever atomic propositions change, even if the investigated system remains the same. 

STL specifications can be combined with statistical verification methods, such as conformal prediction \cite{Vovk2005} and scenario optimization \cite{Campi2018}, to compute probabilistic guarantees on the specification robustness bounds. However, this specification-level approach requires re-sampling when specifications change.
Consequently, online evaluation is computationally challenging 
in practice.
To improve modularity, recent work 
evaluates worst-case atomic robustness over ball-shaped probabilistic reachable sets, using standard STL to compute the specification robustness \cite{Lindemann2023a}. However, this approach cannot be used with other set representations.

In this paper, 
we introduce pacSTL, a probabilistic STL language that evaluates specifications using reachable sets with
Probably Approximately Correct (PAC) guarantees \cite{pacbound}. 
We show that pacSTL transfers the PAC guarantees of the reachable sets to STL specifications, 
enabling efficient evaluation of these specifications.
pacSTL is representation-agnostic, as the relationship between reachable sets and atomic robustness values is enforced through optimization problems rather than specific set representations. 

Compared with other probabilistic STL formulations, our approach avoids explicitly modeling probability distributions over atomic robustness metrics.
This modularity of pacSTL enables more sample- and time-efficient computations than direct uncertainty quantification of the specification robustness, especially when the systems contain rotational and translational invariances. Our main contributions include:
\begin{itemize}
    \item We develop pacSTL to decompose verification under unbounded uncertainty into data-driven reachability and temporal logic evaluation over reachable sets.
    \item We achieve efficient pacSTL computations when specifications change or systems exhibit invariance.
    \item We demonstrate pacSTL on two systems illustrating the computational efficiency and three-valued semantics.
\end{itemize}

\begin{figure*}[h]
    \vspace{0.2cm}
   \centering
   \resizebox{.95\linewidth}{!}{
    \input{Figures/headfigure}
    }
    \caption{Proposed \estl{} framework that combines PAC-bounded reachable tubes with optimization problems that compute lower and upper bounds on atomic robustness, yielding robustness intervals and probabilistic guarantees for \estl{} specifications.}
    \label{fig:pipeline}
\end{figure*}

\section{Preliminaries}
\label{sec:prelininaries}

We denote sets with calligraphic letters, probability distributions over a random variable $x$ as $\mu_x$, and probabilities by~$P$. Further, $\mathbb{I}\mathbb{R}^n$ is the space of intervals in $\mathbb{R}^n$, and interval-valued functions are denoted by $[ f (\cdot) ]$.
Reachable tubes and reachable sets for specific time instants are denoted as $\mathcal{R}$ and $\mathcal{R}_t$, respectively.
Trajectories are denoted by $\delta$, with $\delta_t$ representing the state at time $t$. 
Due to the independence of samples $\delta^{(i)}$, $P^M$ denotes the product probability over $M$ samples, $i=1,\dots,M$.
Additionally, probabilistic bounds are presented with accuracies $\epsilon_{\mathcal{R}}$ and $\epsilon_{\mathcal{R}_t}$ for tubes and time-point sets, respectively, and confidence $\beta$.

\subsection{Data-Driven Reachability Analysis}
\label{DDRA}
A forward reachable set is defined as $\mathcal{R}_t = \{\Xi(t;t_0, x_0, d) : x_0 \in \mathcal{X}_0, d \in \mathcal{D}\}$ where $\mathcal{X}_0 \subseteq \mathbb{R}^{n_x}$ is the set of i.i.d. initial states, $\mathcal{D}$ is a set of i.i.d. disturbance signals $d: [t_0, t] \rightarrow \mathbb{R}^{n_d}$, and $\Xi(t;t_0, \cdot, \cdot) : 
\mathcal{X}_0 \times \mathcal{D} \rightarrow \mathbb{R}^{n_x}$ is the state transition function.
No structural assumptions (e.g., compactness, convexity, etc.) are imposed on $\mathcal{D}$.
$\mathcal{R}_t $ contains all states that the system can transition to at time $t$ from $\mathcal{X}_0$ at $t_0$, subject to disturbances in $\mathcal{D}$.
Further, we define a forward reachable tube as a collection of reachable sets $\mathcal{R} = \{\mathcal{R}_0, \dots, \mathcal{R}_T \}$. We assume that we are given $\mathcal{R}$ with PAC guarantee $ P^M (P_{\delta \in \mathcal{R}} \geq 1 - \epsilon_\mathcal{R} ) \geq 1 - \beta$. Common approaches for computing 
PAC-certified data-driven reachable sets include scenario optimization \cite{Devonport2020, pmlr-v242-dietrich24a}, sample-compression methods \cite{paccagnan2025picktolearnsystemscontroldatadriven, Paccagnan2025-CDC}, and PAC-Bayes approaches \cite{devonport2023}.
\begin{assumption}[PAC-certified reachable sets]
\label{thm:pac}
Given $\beta \in (0, 1)$, reachable tube estimate $\mathcal{R}$, and time-point reachable set estimate $\mathcal{R}_t$, there exists a valid verification procedure that yields PAC guarantees from $M$ finitely many i.i.d. samples satisfying: 
\begin{equation}
\label{equ:probviolatetube}
\begin{aligned}
 P^M ( P_{\delta \in \mathcal{R}} \geq 1 - \epsilon_\mathcal{R} ) &\geq 1 - \beta,
 \end{aligned}
 \end{equation}
 \begin{equation}
 \label{eq:probviolateset} 
 \begin{aligned}
 P^M ( P_{\delta_t \in \mathcal{R}_t} \geq 1 - \epsilon_{\mathcal{R}_t}) &\geq 1 - \beta,
\end{aligned}
\end{equation} 
where $\epsilon_\mathcal{R}$, and $\epsilon_{\mathcal{R}_t}$ depend on the 
verification approach. 
\end{assumption}

\subsection{Interval Signal Temporal Logic (I-STL)}
STL is a formal language to define specifications over time-varying signals \cite{Fainekos2009, maler2004-stllanguage}.
I-STL \cite{Baird2023} extends STL to incorporate bounded uncertainty in signal values and predicate functions. I-STL syntax is the same as STL, with replacement of atomic propositions as interval inclusion functions and Boolean and temporal operators defined over intervals. Specifically, $\mathcal{I}$ is the set of interval inclusion functions where each $\mathcal{M} \in \mathcal{I}$ is an interval function $\mathcal{M}: \mathbb{I}\mathbb{R}^n \rightarrow \mathbb{I}\mathbb{R}$. 
Here, $\mathcal{M}$ provides robustness bounds on atomic propositions for interval bounded signals in $\mathbb{R}^n$. Then, I-STL specifications are formed using the syntax \cite{Baird2023}:
\begin{equation} \label{eq:I-STL_syntax}
    \phi \triangleq (\mathcal{M}([x]) \subseteq [0, \infty])|\neg \phi | \phi \wedge \psi | \phi \,  U_{[t_1, t_2]}\psi,
\end{equation}
where $x$ is a signal, and $\phi \, U_{[t_1, t_2]}\psi$ evaluates to true if 
$\phi$ holds until specification $\psi$ holds on the interval $[t_1, t_2]$.

Most temporal languages, including STL \cite{Fainekos2009}, provide quantitative semantics that characterize the distance to or from satisfaction. This quantitative value of a specification is called \emph{robustness} and is greater than zero when indicating satisfaction.
Similarly, I-STL has quantitative semantics that are constructed from interval robustness functions $\mathbf{h}: \mathbb{I}\mathbb{R}^n \rightarrow \mathbb{I}\mathbb{R}$. For example, the negation $\neg$, disjunction $\vee$, and globally $G_{[t_1, t_2]}$ operations are defined as:
\begin{align}
    \mathbf{h}^{\neg\phi}([x(t)]) &= - \mathbf{h}^{\phi}([x(t)]) \\
    \mathbf{h}^{\phi\vee\psi}([x(t)]) &= [\max](\mathbf{h}^{\phi}([x(t)]), \mathbf{h}^{\psi}([x (t)])) \\
    \mathbf{h}^{G_{[t_1, t_2]}\phi}([x (t)])& =\underset{t' \in [t+t_1, t+t_2]}{[\min]}(\mathbf{h}^{\phi}([x(t')])),
\end{align}
where specifications are denoted by superscripts.
The subscript of the temporal operators, i.e., globally $G$ and eventually $F$, specify the time horizon (when omitted, the time horizon is $[0, T]$, with final time $T$). The interval inclusion functions for $\min$ and $\max$ are denoted by $[\min]$ and $[\max]$. For example, given two arguments,  $[\min]$ is:
\begin{equation}
    [\min] ([x_1],[x_2]) = [\min(\underline{x_1},\underline{x_2}), \min(\overline{x_1},\overline{x_2})] .
\end{equation}
Due to the recursive definition of I-STL \eqref{eq:I-STL_syntax},
the signal intervals $[x]$ become robustness intervals once a Boolean or temporal operation of \ref{eq:I-STL_syntax} is applied. 
The robustness intervals may span both negative and positive values, yielding three-valued semantics \cite[Def. 4]{Baird2023}:
\begin{equation}
    [x(t)] \models \phi =
    \begin{cases}
        \mathtt{True} &\text{if} \, \mathbf{h}^{\phi}([x(t)]) \subseteq [0, \infty]\\
        \mathtt{False} &\text{if} \, \mathbf{h}^{\phi}([x(t)]) \subseteq [-\infty, 0)\\
        \mathtt{Undefined} &\text{otherwise}.
    \end{cases}
\end{equation}
See \cite{Baird2023} for details on the quantitative semantics of I-STL.

\section{PAC-Bounded Signal Temporal Logic (\estl{})}
\label{sec:pacstl}
The 
\estl{} 
language 
provides probabilistic robustness ranges based on PAC-bounded reachable sets (see Fig.~\ref{fig:pipeline}). In this section, we bound atomic propositions over reachable sets and 
relate these bounds to the associated probabilistic guarantees. We then show how to compute {\estl{}} for specifications and derive guarantees on specification satisfaction.

First,  we establish time-point probabilistic guarantees on the robustness bounds of \estl{} atomic propositions. 
\begin{lemma}\label{lemma:atomic_robustness_guarnatuees}
    Given a compact set $\mathcal{R}_t$ with PAC guarantees \eqref{eq:probviolateset}, and the continuous robustness function of an atomic proposition $h(x)$,
    $P^M ( P_{h(\delta_t) \in [\underline{\mathrm{h}}_t, \overline{\mathrm{h}}_t ]} \geq 1- \epsilon_{\mathcal{R}_t}) \geq 1- \beta$, holds, where $\underline{\mathrm{h}}_t \text{ and } \overline{\mathrm{h}}_t$ result from the optimization problems: 
    \begin{equation}
    \label{eq:lowbound}
        \underline{\mathrm{h}}_t = \underset{x}{\mathrm{minimize}} \;  h (x) \; \; \mathrm{subject \; to} \; x \in \mathcal{R}_t,
    \end{equation}
    \begin{equation}
    \label{eq:upbound}
        \overline{\mathrm{h}}_t = \underset{x}{\mathrm{maximize}} \;  h (x) \; \; \mathrm{subject \; to} \; x \in \mathcal{R}_t.
    \end{equation}

    \begin{proof}
The probability that an unseen scenario $\delta_t$ has a robustness within the robustness interval $[\underline{\mathrm{h}}_t, \overline{\mathrm{h}}_t]$ is greater than or equal to the probability that it is contained in the reachable set: $P_{h(\delta_t) \in [\underline{\mathrm{h}}_t, \overline{\mathrm{h}}_t ]} \geq  P_ {\delta_t \in \mathcal{R}_t} $. Every scenario that is inside of the reachable set estimate $\mathcal{R}_t$ has a robustness that lies within $[\underline{\mathrm{h}}_t, \overline{\mathrm{h}}_t]$. Thus, the probabilistic guarantee on the reachable set estimate given by Assumption~\ref{thm:pac} Eq. \eqref{eq:probviolateset} is a conservative guarantee on the robustness:
        \begin{align}
            &P^M (P_ {\delta_t \in \mathcal{R}_t} \geq 1- \epsilon_{\mathcal{R}_t}) \geq 1- \beta \notag \\
            \implies \quad  &P^M (P_{h(\delta_t) \in [\underline{\mathrm{h}}_t, \overline{\mathrm{h}}_t ]} \geq 1- \epsilon_{\mathcal{R}_t}) \geq 1- \beta. \label{eq:timepoint_prob_rob}
        \end{align}
    \end{proof}
 
\end{lemma}

\textbf{Remark 1.} 
The probabilistic guarantee in \eqref{eq:timepoint_prob_rob} can also be computed using statistical verification methods directly applied to the atomic robustness space, i.e., estimating the robustness interval instead of the reachable set. 
However, this is impractical when the robustness functions vary based on parameters, as demonstrated in Sec.~\ref{sec:results}.

\textbf{Example:} 
Fig. \ref{fig:robustness_example} illustrates 
Lemma \ref{lemma:atomic_robustness_guarnatuees} 
for robustness function $h(x) = a^\top x + b, \|a\|_2 =1$, which gives the signed distance from $x$ to the hyperplane $\{y | a^\top y + b=0\}$. The 
ellipsoids represent the PAC-bounded reachable sets over time; the lines indicate where the linear robustness functions are zero (with time-varying $a, b$).
The solutions to \eqref{eq:lowbound} and \eqref{eq:upbound} are determined by  points closest to and farthest from the line. If the line intersects the reachable set, the two farthest points are identified, and their signs are determined by which side of the line they fall. 
In Fig. \ref{fig:robustness_example}, the atomic proposition is robustly satisfied at $t_1$ and falsified at $t_3$. At $t_2$, a negative lower and positive upper bound is calculated---the undefined case in the three-valued semantics \cite{Baird2023}. 

\begin{figure}[h!]
   \centering
    \usetikzlibrary{arrows.meta, positioning, shapes.geometric}

\begin{tikzpicture}[box/.style = {draw, rectangle, minimum width=3cm, minimum height=1cm, align=center}]

\definecolor{darkred}{rgb}{0.55, 0.0, 0.0}
\definecolor{darkcyan}{rgb}{0.0, 0.55, 0.55}
\definecolor{deepcarrotorange}{rgb}{0.91, 0.41, 0.17}
\definecolor{navyblue}{rgb}{0.0, 0.0, 0.5}
\definecolor{fandango}{rgb}{0.71, 0.2, 0.54}
\definecolor{rose}{rgb}{1.0, 0.0, 0.5}

\draw [] (0,0.5) rectangle (8,3.7);
\node[ellipse, draw, minimum width=12mm, minimum height=6mm, color=deepcarrotorange, fill=deepcarrotorange!20] (e) at (1.5,1) {};
\draw[color=darkred, draw opacity=0.5, very thick](0, 3.0) -- (3.333, 0.5); 
\draw[thick, |-|, color=darkred] (2.02,1.16) -- (2.18, 1.355);
\draw (1, 2) node  [font=\scriptsize, color=darkred] [] {$+$};
\draw (1.4, 2.15) node  [font=\scriptsize, color=darkred] [] {$-$};
\draw (2.3, 1.1) node  [font=\scriptsize, color=darkred] [] {$\mathbf{\underline{h}}_1$};
\draw (2.2, 0.7) node  [font=\scriptsize, color=darkred] [] {$\mathbf{t_1}$};
\draw[thick, |-|, color=darkred] (1.107,.78025) -- (1.675, 1.73);
\draw (1.25, 1.5) node  [font=\scriptsize, color=darkred] [] {$\mathbf{\overline{h}}_1$};

\node[ellipse, draw, minimum width=18mm, minimum height=9mm, color=darkcyan, fill=darkcyan!20] (e) at (3.8,1.8) {};
\draw[color=navyblue, draw opacity=0.5, very thick](6.625, 3.7) -- (2.625, 0.5);
\draw[thick, |-|, color=navyblue] (4.545,1.555) -- (4.27, 1.84);
\draw (4.55, 2.3) node  [font=\scriptsize, color=navyblue] [] {$+$};
\draw (4.85, 2.05) node  [font=\scriptsize, color=navyblue] [] {$-$};
\draw (4.55, 1.3) node  [font=\scriptsize, color=navyblue] [] {$\mathbf{\underline{h}}_{2}$};
\draw (3.95, 1.15) node  [font=\scriptsize, color=navyblue] [] {$\mathbf{t_2}$};
\draw[thick, |-|, color=navyblue] (3.72, 1.355) -- (3.125, 2.11);
\draw (3.05, 2.3) node  [font=\scriptsize, color=navyblue] [] {$\mathbf{\overline{h}}_2$};

\node[ellipse, draw, minimum width=24mm, minimum height=12mm, color=fandango, fill=fandango!20] (e) at (6.5,3.0) {};
\draw[color=rose, draw opacity=0.5, very thick](2.452, 3.7) -- (3.012, 0.5);
\draw (2.7, 3.45) node  [font=\scriptsize, color=rose] [] {$-$};
\draw (2.3, 3.45) node  [font=\scriptsize, color=rose] [] {$+$};
\draw (6.55, 2.2) node  [font=\scriptsize, color=rose] [] {$\mathbf{t_3}$};
\draw[thick, |-|, color=rose] (7.7,3.05) -- (2.6, 2.75);
\draw (4.5, 3.25) node  [font=\scriptsize, color=rose] [] {$\mathbf{\overline{h}}_3$};
\draw[thick, |-|, color=rose] (5.3,3.1) -- (2.57, 2.95);
\draw (6, 2.75) node  [font=\scriptsize, color=rose] [] {$\mathbf{\underline{h}}_3$};

\draw (7.5, 1) node  [font=\normalsize, color=black] [] {$\mathcal{X}$};
\end{tikzpicture}
    \caption{
    Robustness bounds 
    from PAC-bounded ellipsoidal reachable sets for three time steps and varying atomic robustness functions in state space $\mathcal{X}$. The lines are where the robustness functions equal zero, i.e., $h_i(x)=a_i^\top x + b _i = 0$. The colors indicate different time steps, while $+$ and $-$ denote which side corresponds to satisfaction and falsification, respectively.
   }
\label{fig:robustness_example}
\end{figure}

The solutions to \eqref{eq:lowbound} and \eqref{eq:upbound} are the global minimum and maximum of $h(x)$ where $x \in \mathcal{R}_t$. Therefore, these two optimization problems result in an inclusion function for $h$ as 
in \cite[Def. 5]{Baird2023}. It then follows from \cite[Thm. 1]{Baird2023} that the quantitative I-STL semantics are sound:
\begin{proposition}[Soundness of pacSTL---adapted from \cite{Baird2023}]
\label{prop:semanticlink}
The solution to \eqref{eq:lowbound} and \eqref{eq:upbound} is an atomic proposition interval extension, which is an interval inclusion function \cite[Def. 5]{Baird2023}, and the induced specification-level robustness is a natural inclusion function \cite[Prop. 2]{Baird2023}. Thus, if the specification is satisfied for the interval induced by \eqref{eq:lowbound} and \eqref{eq:upbound}, it also holds for any signal within the interval bound.
\end{proposition}

STL specifications are defined over a trace; therefore, we build on Lemma \ref{lemma:atomic_robustness_guarnatuees} and Proposition~\ref{prop:semanticlink} to define \estl{} over a reachable tube to evaluate an STL specification $\phi$.

\begin{theorem}[Reachable Tube \estl] \label{thm:tube}
Let the robustness intervals for atomic predicates be computed according to \eqref{eq:lowbound} and \eqref{eq:upbound} with respect to the time-point reachable set estimates of the reachable tube, let the probabilistic guarantee for the reachable tube be defined as in Assumption~\ref{thm:pac} Eq. \eqref{equ:probviolatetube}, and let the robustness interval $[\underline{\mathrm{h}}^\phi, \overline{\mathrm{h}}^\phi]$ for the STL specification $\phi$ be calculated based on I-STL quantitative semantics  \cite[Def. 3]{Baird2023}. Then, the following probabilistic guarantee holds: 
\begin{equation} \label{eq:thm2_bound}
     P^M( P_{h^\phi(\delta) \in [\underline{\mathrm{h}}^\phi, \overline{\mathrm{h}}^\phi ]} \geq 1- \epsilon_{\mathcal{R}}) \geq 1- \beta.
\end{equation}

\begin{proof}
Given Proposition~\ref{prop:semanticlink}, $[\underline{\mathrm{h}}^\phi, \overline{\mathrm{h}}^\phi]$ is sound when computed based on $[\underline{\mathrm{h}}_t^i, \overline{\mathrm{h}}_t^i]$ where $i \in \{1, ..., K\}$, $t \in \{0, ..., T\}$, and $K$ is the number of atomic propositions. 
As in Lemma~\ref{lemma:atomic_robustness_guarnatuees}, if a trajectory is fully contained in the reachable tube $\mathcal{R}$, then its robustness with respect to specification $\phi$ will be within $[\underline{\mathrm{h}}^\phi, \overline{\mathrm{h}}^\phi]$ due to the conservative calculation of the atomic robustness intervals.
Thus, $  P_{h^\phi(\delta) \in [\underline{\mathrm{h}}^\phi, \overline{\mathrm{h}}^\phi]} \geq  P_{\delta \in \mathcal{R}}$, and from Assumption~\ref{thm:pac} Eq. \eqref{equ:probviolatetube} we obtain:
\begin{align}
    &P^M (P_{\delta \in \mathcal{R}}\geq 1- \epsilon_{\mathcal{R}}) \geq 1- \beta \notag \\
    \implies \quad & P^M (P_{h^\phi(\delta) \in [\underline{\mathrm{h}}^\phi, \overline{\mathrm{h}}^\phi]} \geq 1- \epsilon_{\mathcal{R}}) \geq 1- \beta.
\end{align}
\end{proof}
\end{theorem}

\textbf{Practical Implementation.}
The key requirement for using \estl{} is the availability of atomic robustness functions and reachable set representations that enable exact computation of \eqref{eq:lowbound} and \eqref{eq:upbound}. This requirement is satisfied for linear and quadratic robustness functions over convex sets. The restriction on atomic robustness functions is mild, as linear and quadratic functions cover most common atomic propositions.
Nonlinear functions can be handled by augmenting the state space to include the nonlinear output or by developing specialized algorithms, which is often feasible for trigonometric functions. 
The \estl{} framework can also use exact or over-approximated model-based reachable sets, resulting in robustness intervals that hold deterministically. When PAC-bounded reachable sets are used, it is assumed that the underlying distribution of the initial states and disturbances does not change between reachable set construction and deployment.
Additionally, \estl{} can be used for continuous time with model-based time-interval reachable sets, as the formulation is not restricted to discrete time. However, to the best of our knowledge, there currently exist no data-driven methods for computing continuous-time, time-interval reachable sets with PAC guarantees solely from samples.
Lastly, note that the formulation of \eqref{eq:lowbound} and \eqref{eq:upbound} is the main factor affecting the runtime of online evaluation of \estl{} specifications.

\begin{table}[t]
\vspace{.2cm}
\centering
\renewcommand{\arraystretch}{1.5}
\caption{Accuracies of Reachable Sets}
\label{tab:epsilons}
\begin{tabular}{lccc}
\toprule
\textbf{System} & $\epsilon_{\mathcal{R}}$ & $\underset{t}{\min} \hspace{1mm} \epsilon_{\mathcal{R}_t}$ & $\underset{t}{\max} \hspace{1mm}\epsilon_{\mathcal{R}_t}$ \\ \midrule
Quadrotor       &  0.0912   &  0.0533  &  0.0846 
\\ \midrule
Vessel   &  0.0412 & 0.0263 & 0.0346 \\ \bottomrule
\end{tabular}
\vspace{-.3cm}
\end{table}

\section{Numerical Experiments}\label{sec:results}

We demonstrate \estl{} through two case studies: verification of quadrotor control \cite{Meyer2021} and online monitoring of vessel collision-avoidance.\footnote{For reproducibility, we provide our Python implementation including parameters at: \url{https://github.com/eadietri/pacSTL_LCSS}}
We use methods that provide PAC guarantees without requiring distributional assumptions. We do not compare to conformal prediction methods \cite{lindemann2025formalverificationcontrolconformal}, as existing reachability-analysis formulations generally only provide marginal coverage guarantees.
All numerical experiments were performed on a machine with an Intel Core i7-12800H CPU and 32 GB of RAM, using SciPy 
\cite{virtanen2020scipy} 
to solve \eqref{eq:lowbound} and \eqref{eq:upbound}. We first describe the computation of reachable tubes, and then report \estl{} results for the two systems.  
Note that our experiments have no distribution shift from reachable set calculation to deployment of the to-be-verified system.

\vspace{-3mm}
\subsection{Data-Driven Reachable Sets}

While there are multiple approaches to obtain PAC-bounded reachable sets \cite{Devonport2020, pmlr-v242-dietrich24a, paccagnan2025picktolearnsystemscontroldatadriven, Paccagnan2025-CDC,devonport2023}, we employ sampling-based optimization to obtain ellipsoids \cite{Devonport2020} and perform verification using the holdout method \cite{dietrich2025datadrivenreachabilityscenariooptimization}.
At each time step, we solve the program \cite[Eq. (8)]{Devonport2020} with CVXPY  \cite{diamond2016cvxpy} 
to obtain ellipsoidal, time-point reachable sets.
We use $N=1500$ i.i.d. training samples to generate the reachable sets and $M=1500$ i.i.d. testing samples with $\beta=10^{-9}$ to compute probabilistic guarantees. 
Tab.~\ref{tab:epsilons} reports the accuracies of the reachable tubes and the minimum and maximum accuracies of the time-point reachable sets, empirically confirming that time-point PAC bounds are tighter.

\vspace{-3mm}
\subsection{Quadrotor}
We augment the 12-dimensional PD-controlled system described in \cite[Chap. 8.2]{Meyer2021} with disturbances and initial conditions sampled from Normal distributions to mimic unbounded real-world uncertainty. 
We aim to verify that the quadrotor never exceeds the height $p_z$ of $\SI{1.5}{\meter}$, stays above $\SI{0.85}{\meter}$ meters after the first second, and exhibits low angular or translational velocities. Formally:
\begin{align*}
    \Phi: \; &F_{[0, 4 \Delta t]} G (p_z \geq 0.85) \land G(p_z \leq 1.5) \land \\
    &G(\|[\dot{p_n},\dot{p_e},\dot{p_z}]\|_2^2 \leq 1 \lor \|[\dot{\phi},\dot{\theta},\dot{\psi}]\|_2^2 \leq 1).
\end{align*}
The four atomic propositions are either linear or quadratic in the state dimension, and the time horizon is $T=\SI{5}{\second}$ with sampling period $\Delta t = \SI{0.25}{\second}$.
To investigate the conservativeness introduced by using reachable sets, we compute a sample-inefficient PAC guarantee directly on the robustness, called the direct approach. Specifically, we compute the robustness interval from $N=1500$ trajectories and evaluate on an additional $M=1500$ test samples. 
We use the number of test samples with robustness values outside the interval to compute $\epsilon$ (\cite[Def. 1]{dietrich2025datadrivenreachabilityscenariooptimization}) for the given $M$ and $\beta = 10^{-9}$.

\begin{figure}
\vspace{0.2cm}
    \centering
        \definecolor{CaliforniaGold}{rgb}{0.992, 0.71, 0.082}
\definecolor{BerkeleyBlue}{rgb}{0.0, 0.149, 0.463}
\definecolor{darkblue}{rgb}{0.0, 0.13, 0.28}
\definecolor{darkred}{rgb}{0.55, 0.0, 0.0}
\definecolor{darkcyan}{rgb}{0.0, 0.55, 0.55}
\definecolor{deepcarrotorange}{rgb}{0.91, 0.41, 0.17}
\definecolor{navyblue}{rgb}{0.0, 0.0, 0.5}
\definecolor{fandango}{rgb}{0.71, 0.2, 0.54}
\definecolor{rose}{rgb}{1.0, 0.0, 0.5}

\begin{tikzpicture}
        \begin{axis}[
            width=.45\textwidth,
            height=4cm,
            xmin=0, xmax=5,
            ymin=-1, ymax=2,
            xlabel={$t$ $(\unit{\second})$},
            ylabel={$p_z$ $(\unit{\meter})$},
            grid=major,
        ]

        \addplot[
            name path=line,
            darkred,
            thick,
            domain=-1:6
        ] {1.5};
        
        \addplot[
            name path=top,
            draw=none,
            domain=-1:6
        ] {2.75};
        
        \addplot[
            darkred,
            opacity=0.3
        ] fill between[of=line and top];

        \draw [draw=darkred, thick, fill=darkred, fill opacity=0.3] (1, -1.75) rectangle (6, 0.85);

        \addplot[
            BerkeleyBlue,
            thick,
        ] coordinates {
            (0.25, -0.72)
            (0.25, 0.91)
        };

         interval [ymin, ymax] 
        \addplot[
            BerkeleyBlue,
            thick,
        ] coordinates {
            (0.5, -0.183212)
            (0.5, 1.1032796)
        };

        \addplot[
            BerkeleyBlue,
            thick,
        ] coordinates {
            (0.75, 0.503)
            (0.75, 1.2067)
        };
        
        \addplot[
            BerkeleyBlue,
            thick,
        ] coordinates {
            (1.0, 0.989)
            (1.0, 1.279)
        };

        \addplot[
            BerkeleyBlue,
            thick,
        ] coordinates {
            (1.5, 1.0276)
            (1.5, 1.449)
        };

        \addplot[
            BerkeleyBlue,
            thick,
        ] coordinates {
            (2.0, 0.9498)
            (2.0, 1.1477)
        };

        \addplot[
            BerkeleyBlue,
            thick,
        ] coordinates {
            (2.5, 0.8951)
            (2.5, 0.98052)
        };

        \addplot[
            BerkeleyBlue,
            thick,
        ] coordinates {
            (3, 0.9137)
            (3, 1.0038)
        };

        \addplot[
            BerkeleyBlue,
            thick,
        ] coordinates {
            (3.5, 0.993722)
            (3.5, 1.01595)
        };

        \addplot[
            BerkeleyBlue,
            thick,
        ] coordinates {
            (4, 1.0032)
            (4, 1.0286)
        };
        
        \addplot[
            BerkeleyBlue,
            thick,
        ] coordinates {
            (4.5, 0.9976)
            (4.5, 1.0115)
        };
        
        \addplot[
            BerkeleyBlue,
            thick,
        ] coordinates {
            (4.75, 0.9962)
            (4.75, 1.0023)
        };


        \addplot[
            name path=upper,
            draw=none
        ] coordinates {
            (0.0, 0.3)
            (0.25, 0.91)
            (0.5, 1.10328)
            (0.75, 1.2067)
            (1.0, 1.279)
            (1.25, 1.4320620)
            (1.5, 1.449)
            (1.75,  1.321)
            (2.0, 1.1477)
            (2.25,  1.0131680)
            (2.5, 0.98052)
            (2.75,  0.9909)
            (3.0, 1.0038)
            (3.25, 1.01173)
            (3.5, 1.01595)
            (3.75, 1.0251)
            (4.0, 1.0286)
            (4.25, 1.0220)
            (4.5, 1.0115)
            (4.75, 1.0023)
        };
        
        \addplot[
            name path=lower,
            draw=none
        ] coordinates {
            (0.0, -0.3)
            (0.25, -0.72)
            (0.5, -0.1832)
            (0.75, 0.503)
            (1.0, 0.989)
            (1.25, 1.0737)
            (1.5, 1.0276)
            (1.75, 0.97895)
            (2.0, 0.9498)
            (2.25, 0.93303)
            (2.5, 0.8951)
            (2.75, 0.88501)
            (3.0, 0.9137)
            (3.25, 0.95754)
            (3.5, 0.993722)
            (3.75, 1.0048)
            (4.0, 1.0032)
            (4.25, 1.0000)
            (4.5, 0.9976)
            (4.75, 0.9962)
        };

        \addplot[
            BerkeleyBlue,
            opacity=0.25
        ] fill between[of=upper and lower];



        \end{axis}
\end{tikzpicture}
        \vspace{-0.2cm}
    \caption{Verification of height components of quadrotor specification using reachable sets projected to $p_z$. The time-point sets are highlighted in dark blue for every time point. 
    The set of initialized heights is approximated by two standard deviations of the sampled Normal distribution.
    }
\label{fig:quadrotorspecsat}
    
\end{figure}

We verify the specification with \estl{} and the direct approach, resulting in robustness intervals $[0.0350, 0.1305]$ and $[0.0631, 0.1195]$, and accuracies $0.0912$ (Tab.~\ref{tab:epsilons}) and $0.0222$, respectively. The average runtime of the \estl{} evaluation over ten runs is $\SI{0.4}{\second}$, while the direct approach on average takes $\SI{6.6}{\second}$, where the majority of the runtime stems from the sample generation ($\SI{6.1}{\second}$).
Fig. \ref{fig:quadrotorspecsat} illustrates the satisfaction of the height-related components of the specification.
If the specification changes, e.g., the height exceeds 0.95 meters, we do not require new samples to re-compute robustness intervals with \estl{}.

\vspace{-3mm}
\subsection{Vessel Navigation}
To demonstrate runtime verification with \estl{}, we investigate a head-on encounter in two-vessel navigation with a specification simplified from maritime traffic rules \cite{Krasowski2021.MarineTrafficRules}. We consider a westbound, ego vessel that predicts its future trajectory and is controlled by a waypoint-tracking controller, which generates evasive waypoints when the specification condition is met. The trajectory of the obstacle vessel is simulated with a six-degree-of-freedom (6-DOF) model (Eqs. (2.1) and (2.2), \cite{fossen2011handbook}), which we also use for computing the data-driven reachable sets. Due to rotational and translational invariance, the offline-computed reachable sets for the obstacle vessel can be reused during operation through appropriate translations and rotations of the scenario, provided that a suitable set of initial vessel velocities is considered during reachable-set computation.

For reachable sets and trajectory prediction, we consider the five dimensions relevant to the specification: surface position $\mathbf{p} = [p_x, p_y]^\top$, orientation $\psi$, and velocity $\mathbf{v} = [v_x, v_y]^\top$.
To account for the shape of the vessel, we transform the center coordinates $\mathbf{p}$ to four data points that represent the corners.
We formulate a specification that evaluates, over prediction horizon $T=\SI{2.5}{\second}$ with discretization $\Delta t = \SI{0.5}{\second}$, whether the obstacle vessel is in front of the ego and whether the relative velocity can result in a collision within time horizon $t_h$:
\begin{align*}
    \Phi : \; &G_{[\Delta t, T]}\big( a(t)^\top \mathbf{p}_o (t) -\, b(t)  \geq 0 \; \land \\
    &\|\mathbf{v}_e (t) - \mathbf{v}_o(t)\|_2^2 - \frac{(\|\mathbf{p}_e (t) - \mathbf{c}_o (t) \|_2 + r_e)^2}{t_h^2}  \geq 0) \big),
\end{align*} \vspace{-0.4cm}
\begin{align*}
    a(t) =\frac{-1}{v_{\mathrm{max}}}\begin{bmatrix} - \sin(\psi_e(t) + \pi/2)  \\ 
    \cos(\psi_e(t) + \pi/2) 
    \end{bmatrix}, \; b(t) = a(t)^\top \,\mathbf{p}_e(t).
\end{align*}
The subscripts $e$ and $o$ indicate states of the ego and obstacle vessel, respectively. Further, $\mathbf{c}_o$ is the center of the reachable set, and $v_{\mathrm{max}}, r_e$, and $t_h$ are fixed parameters.

\addtolength{\textheight}{-5mm}  

Fig.~\ref{fig:robustness_trace_vessel}  displays the robustness bounds over time and vessel traces in the position space for four evaluations. The two red runs are 142 time steps long and use worst-case semantics, i.e., initiating the evasive maneuver once the upper robustness bound becomes positive, while the two blue runs only begin evading once the full robustness interval is positive and have 147 steps. 
This demonstrates different use cases of the three-valued semantics. Additionally, note that the runtime to evaluate \estl{} was on average $\SI{82}{\milli \second}$.

\begin{figure}[tb]
\vspace{0.25cm}
    \centering
    \input{Figures/vessel_results}
    \vspace{-0.2cm}
    \caption{Top: Robustness bounds, with zoom in for $\SI{0}{\second}-\SI{35}{\second}$. Bottom: Position space trajectories for four evaluations. Vessels are displayed along the trajectory of the dark-blue run at two time points. The dark-blue and cyan runs use optimistic decision making, requiring the full robustness interval of the specification to be greater than zero. The pink and maroon runs use worst-case semantics, triggering the collision avoidance maneuver once only the upper bound is positive.
    \vspace{-0.2cm}
    }
    \label{fig:robustness_trace_vessel}
    \vspace{-0.2cm}
\end{figure}

\section{Characteristic-Time Points and PAC-bound}\label{sec:characteristic}
The \estl{} robustness intervals computed from reachable tubes may be conservative due to over-approximation in both the reachable set computation and the interval-valued I-STL semantics.
We investigated this conservativeness by evaluating $150, 000$ samples---100 times the number of samples used to compute the PAC bound---for the quadrotor and for ten vessel states along their trajectories. For all samples, the STL robustness remained within the pacSTL interval $[\underline{\mathrm{h}}^\phi, \overline{\mathrm{h}}^\phi]$, confirming 
the expected conservativeness.

In this section, we discuss a direction for improving the accuracy of pacSTL. First, we extend the quantitative semantics of I-STL to track the time points from which the specification-level robustness interval originates.
\begin{definition} (Characteristic Time Points for I-STL Quantitative Semantics). 
The characteristic time points denote the time points of the smallest lower bound and the largest upper bound over the time interval $[\tau_1, \tau_2]$ of a time-dependent interval-valued function $y(\cdot)$:
\begin{align}
    \tau_{low} (y(\cdot)) &= \underset{\tau ' \in [\tau_1, \tau_2]}{\arg\min} [\min] (y(\tau ')), \\
    \tau_{up}(y(\cdot)) &= \underset{\tau' \in [\tau_1, \tau_2]}{\arg\max} [\max] (y(\tau ')). 
\end{align}
Specifically, for an interval-valued robustness function of the specification $\phi$, we define
\begin{equation}
    \underline{t} = \tau_{low}(\mathbf{h}^\phi), \quad
     \overline{t} = \tau_{up}(\mathbf{h}^\phi).
\end{equation}
\end{definition}
By tracking characteristic time points, the lower and upper \estl{} robustness bounds can be associated with the specific reachable sets from which they arise. We observed empirically for $150, 000$ quadrotor samples, that the characteristic time points of the direct approach's STL evaluation matched those of pacSTL. In the vessel example, this held for at least $85\%$ of the samples for each of the ten states investigated, while for six of the ten states, the characteristic time points of the direct approach's STL evaluation match \estl{} perfectly.

In view of these observations,
assume that the true lower and upper bounds on the robustness $\mathbf{h}^\phi$ originate from the characteristic time points. 
Additionally, 
assume a worst-case evaluation setting where the same $M$ i.i.d. test samples are reused across all time points when computing $\epsilon_{\mathcal{R}_t}$ in Assumption~\ref{thm:pac}. 
Then, from Lemma~\ref{lemma:atomic_robustness_guarnatuees} and Proposition~\ref{prop:semanticlink}, 
\begin{equation}\label{eq:ctp_probbound_lower}
    P_{h^\phi(\delta) \geq \underline{\mathrm{h}}^\phi} \geq P_{\delta_{\underline{t}} \in \mathcal{R}_{\underline{t}}}, \quad  P_{h^\phi(\delta) \leq \overline{\mathrm{h}}^\phi} \geq P_{\delta_{\overline{t}} \in \mathcal{R}_{\overline{t}}} \, .
\end{equation}
To relate these inequalities to \eqref{eq:thm2_bound}, 
note that the robustness interval requires both the upper and lower bounds to hold simultaneously. By applying the lower bound of the Fréchet inequality for the intersection of the events in \eqref{eq:ctp_probbound_lower},
we obtain:
\begin{align}\label{eq:different_times_thm3_proof}
    P_{h^\phi(\delta) \in [\underline{\mathrm{h}}^\phi, \overline{\mathrm{h}}^\phi]}
    &\geq \max \left(P_{\delta_{\underline{t}} \in \mathcal{R}_{\underline{t}}} + P_{\delta_{\overline{t}} \in \mathcal{R}_{\overline{t}}} - 1, 0 \right) \\
    &\geq \max \left((1 - \epsilon_{\mathcal{R}_{\underline{t}}}) + (1 - \epsilon_{\mathcal{R}_{\overline{t}}}) - 1, 0 \right) \notag \\
    &= \max \big(1 - \underbrace{(\epsilon_{\mathcal{R}_{\underline{t}}} + \epsilon_{\mathcal{R}_{\overline{t}}})}_{= \epsilon_{ct}}, 0 \big) \notag 
\end{align}
Since $\epsilon_{ct}$ is small by construction of the reachable sets, $\max$ returns $1 - \epsilon_{ct}$. By similar application of the Fréchet inequality to the outer probability bound, we obtain $\beta_{ct}=2\beta$. With the the two Fréchet inequality applications, and assuming worst-case use of the $M$ i.i.d. testing samples,
it follows that
\begin{align} \label{eq:characteristictimepoint_guarantee}
    P^M ( P_{h^\phi(\delta) \in [\underline{\mathrm{h}}^\phi, \overline{\mathrm{h}}^\phi]} \geq 1- \epsilon_{ct}) \geq 1 - \beta_{ct}. 
\end{align}
For the special case $\underline{t} = \overline{t}$, $\epsilon_{ct} = \epsilon_{\mathcal{R}_{\underline{t}}} = \epsilon_{\mathcal{R}_{\overline{t}}}$ and $\beta_{ct} = \beta$.

For the quadrotor example, the characteristic time points are $\underline{t}=\SI{2.75}{\second}$ and $\overline{t}=\SI{2.5}{\second}$ resulting in $\epsilon_{ct} = 0.0629 + 0.0748 = 0.1377$.
For the vessel, the accuracy $\epsilon_{ct}$ is $0.0364$ on average across all monitoring time steps, with a minimum of $0.0286$ and a maximum of $0.0574$. 
In general, we believe that the notion of characteristic time points is useful to identify conservativeness in the reachable sets.

\section{Conclusion} \label{sec:conclusion}

We propose \estl{}, a STL language based on PAC-bounded reachable sets, which can be used for monitoring and verifying discrete-time black-box systems under unbounded uncertainty. 
Our experiments demonstrate that \estl{} is efficiently computable with ellipsoidal reachable sets and linear or quadratic atomic propositions, enabling runtime monitoring of time-varying specifications.
Future research should integrate \estl{} with control synthesis and develop guarantees that remain valid under distribution shifts.






\bibliography{ref}
\bibliographystyle{IEEEtran}

\end{document}